\documentstyle[sprocl]{article}

\input epsf

\def\Journal#1#2#3#4{{#1} {\bf #2}, #3 (#4)}

\def\PRD{{\em Phys. Rev.} D}

\def\CPC{\em  Comput.\ Phys.\ Commun. }

\def\be{\begin{equation}}
\def\ee{\end{equation}}
\def\bea{\begin{eqnarray}}
\def\eea{\end{eqnarray}}

\def\dofig#1#2{\epsfxsize=#1\centerline{\epsfbox{#2}}}
\def\dofigs#1#2#3{\centerline{\epsfxsize=#1\epsfbox{#2}\ \ \ \ \ \  \epsfxsize=#1\epsfbox{#3}}}

\begin{document}

\title{\mbox{SUSYGEN3, AN EVENT GENERATOR FOR LINEAR COLLIDERS}}

\author{ N. GHODBANE}

\address{Institut de Physique Nucl\'eaire de Lyon \\
43 Bd du 11 novembre 1918,  69622 Villeurbanne cedex, France}

\maketitle\abstracts{ The Monte Carlo program \texttt{SUSYGEN}, 
initially designed for computing distributions and generating
events for supersymmetric particle production 
in $e^+e^-$ collisions, has now been upgraded to study
supersymmetric processes at linear colliders by the inclusion 
of beamstrahlung, beam polarization, spin correlations and
complex couplings including CP violating phases. It
continues to offer, in the new context,  the possibility to study
the production and decay of supersymmetric particles within the 
most general minimal standard supersymmetric model (MSSM), the
minimal supergravity model (MSUGRA) or the gauge mediated
supersymmetry breaking model (GMSB), with or without assuming
R-parity conservation.}

\section{SUSYGEN2 content and upgrades}
Hopefully the linear collider would  operate in an environment 
where  a few sparticles would have been discovered and the high
statistics will permit a precise study of the soft supersymmetry
 breaking parameters \cite{haberandkane}. One has
therefore to move from a \textit{searches} Monte Carlo to a
\textit{"precision"} physics Monte Carlo since one would need to
measure accurately the above parameters.
\texttt{SUSYGEN2} \cite{kats:susygen2} is a Monte Carlo event 
generator for the production and decay of
supersymmetric particles and has been initially 
designed for $e^+e^-$ colliders. It
has been extensively used by all four LEP
experiments to simulate the expected signals.
It includes pair production of charginos and neutralinos, 
scalar leptons and quarks. It offers also the possibility to
study the production of a gravitino plus a neutralino within GMSB models
 and the production of single gauginos if one assumes R-Parity to be broken 
All important decay modes of supersymmetric particles relevant 
to LEP energies have been implemented, including cascades, radiative 
decays and R-Parity violating decays to standard model particles.
The decay of particles is done through the exact matrix 
elements.
The lightest supersymmetric particle (LSP) can either  be the neutralino 
($\tilde{\chi}^0_1$), the sneutrino ($\tilde{\nu}$) or the
gravitino ($\tilde{G}$) in R-Parity conserving models, 
or there can be no LSP if R-Parity is violated.
The initial state radiative corrections  take account of  
$p_T / p_L$ effects  in the Structure Function formalism. 
QED final state radiation is implemented using the \texttt{PHOTOS}
\cite{was:photos} library. 
An optimized hadronization interface to \texttt{JETSET}
\cite{sjos:jetset} is  provided, 
which also takes into account  the lifetime of sparticles. Finally, a widely used feature of \texttt{SUSYGEN2} is the possibility to do automatic scans on the parameter space through user friendly ntuples.
Recently \texttt{SUSYGEN2}  has been upgraded to
\texttt{SUSYGEN3} \cite{nous:susygen3} in order  to adapt to the needs of
the next generation of linear colliders, but also in order to
extend its potential to supersymmetric particles searches 
at $e^-p$ colliders (e.g HERA) and hadronic colliders (e.g
TEVATRON or LHC).
The main new features relevant for linear colliders are
the inclusion of beamstrahlung through an interface to
\texttt{CIRCEE} \cite{ohl:beamstrah}, the full spin correlation in
initial and final states, the inclusion of CP violating phases and
the possibility to have an elaborate calculation of the MSUGRA
spectrum through an interface to \texttt{SUSPECT}
\cite{djoua:suspect}.

\section{Beam polarization and spin correlations}
\begin{figure}[h!]
\vskip -1cm
\dofigs{2.3in}{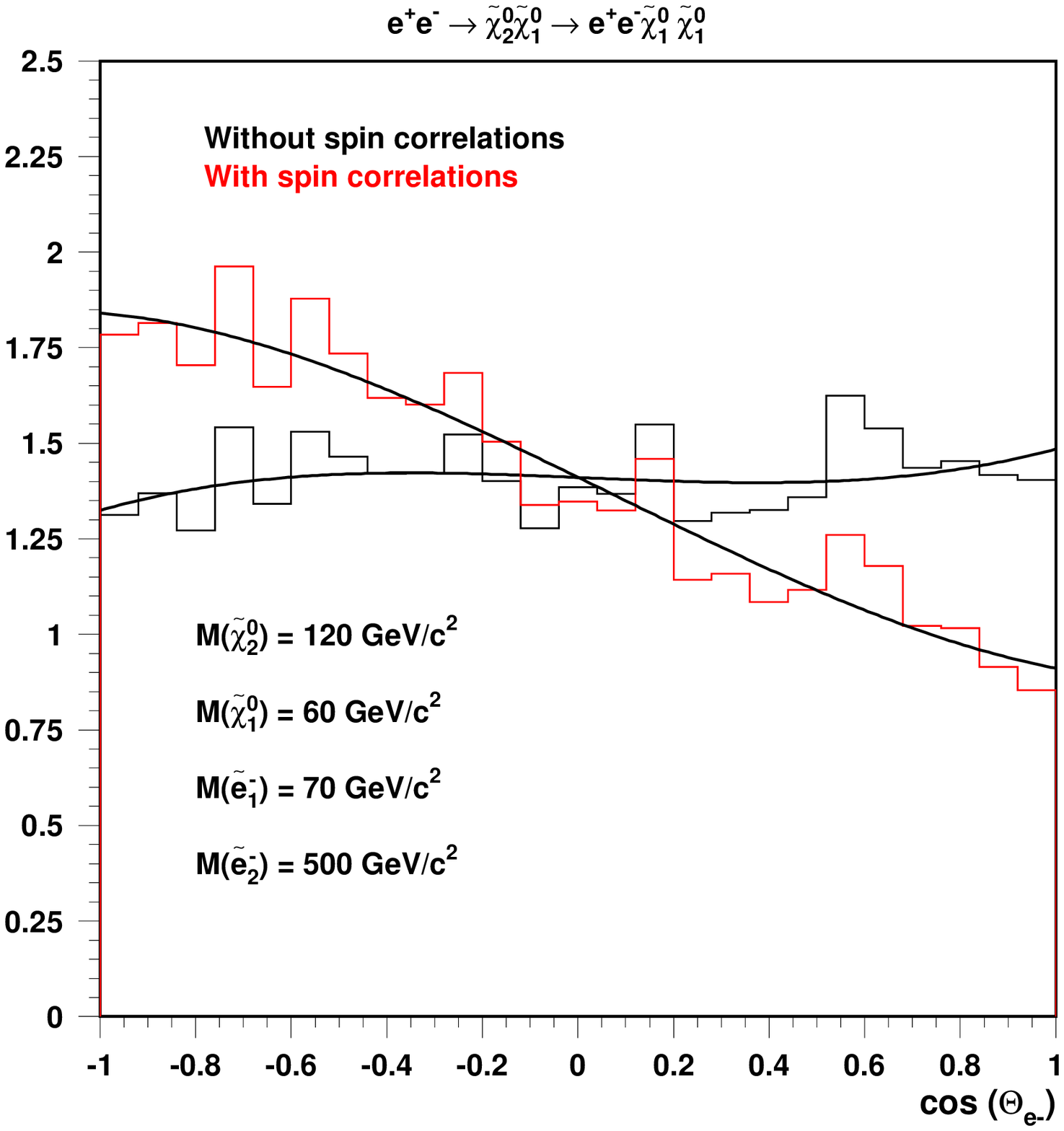}{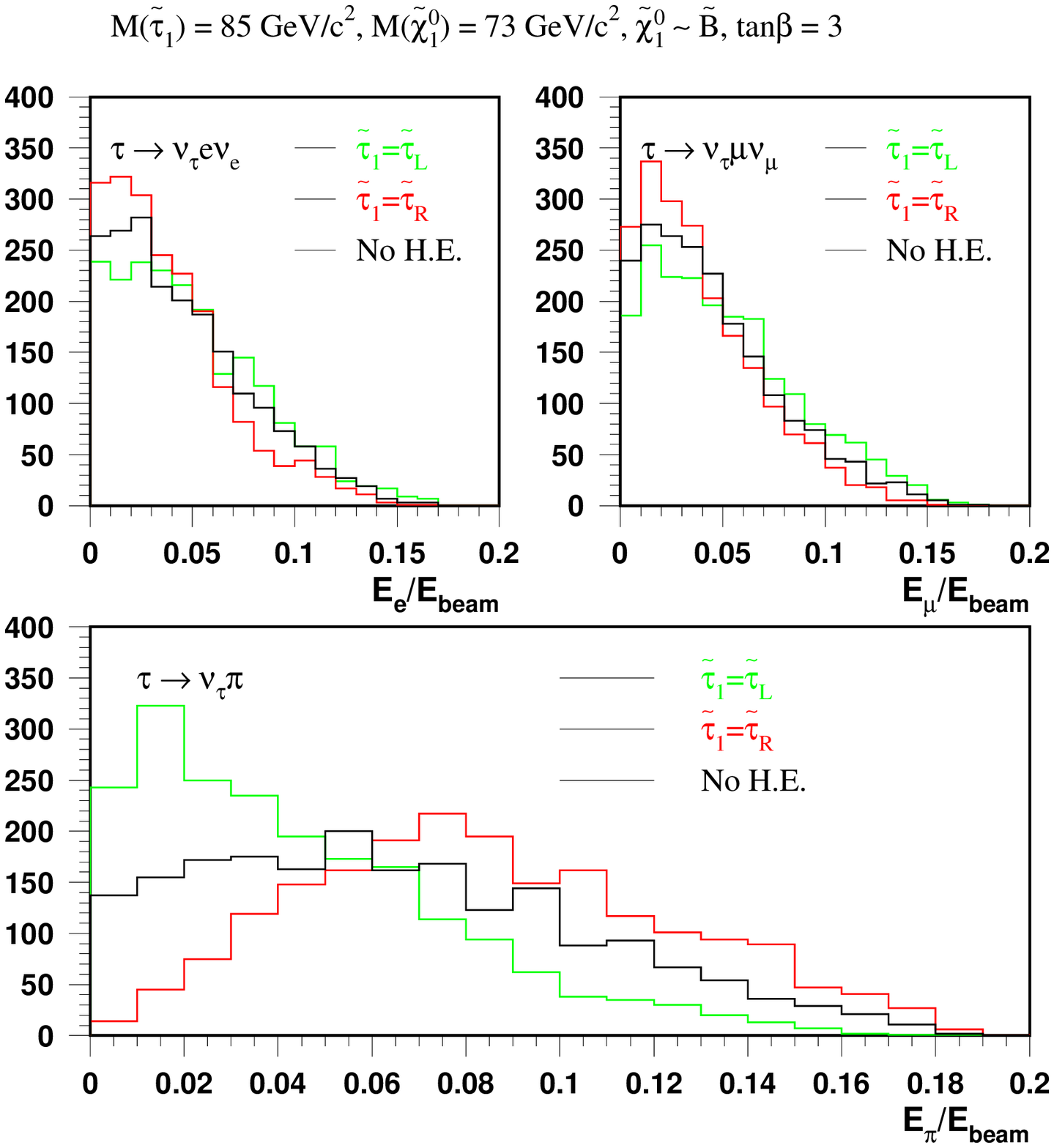}
\vskip -1cm
\caption{ The figure on the left side shows the $d\sigma/d\cos\theta$ distribution for the final $e^-$ associated to the process $e^+e^-\to \tilde{\chi}^0_2\tilde{\chi}^0_1$ with and without taking into account spin correlations. The figure on the right shows the momentum distribution of $e,\mu$ and $\pi$, decay products of $\tau$ produced in the process $e^+e^- \to \tilde{\tau}_1^+ \tilde{\tau}_1^- \to \tau^+ \tilde{\chi}^0_1 \tau^- \tilde{\chi}^0_1$. The distributions have been plotted for two hypotheses concerning the stau mixing angle $\theta_{\tilde{\tau}}$ ($\tilde{\tau}_L$ and $\tilde{\tau}_R$) and compared to the no spin correlation case \label{fig:beampolcorr}}
\end{figure}

Since one expects high luminosities for the next generation of 
linear colliders (e.g. $\sim 500 fb^{-1}$  for the TESLA project),
one can use beam polarization to reduce the standard model 
backgrounds and use the polarization dependence of the cross
sections to study specific SUSY parameters.
Moreover, as it has been stressed by several authors
\cite{gudi:spincorr}, spin correlations play a major role in the
kinematic distributions of final particles.  To fulfill these
 two requirements, the
\textit{helicity amplitude method} \cite{mana:helicity} was used 
for the calculation of the different feynman amplitudes for
production and decay, in order to obtain full spin correlation. 
Since such an amplitude involves products and contractions of
fermionic currents, two basic functions, namely the $B$ and 
$Z$ functions were defined through:
\begin{eqnarray*}
\begin{array}{lll}
B_{\lambda_1,\lambda_2}^{\lambda} (p_1,p_2) &=& \bar{u}_{\lambda_1}(p_1,m_1) 
                                               P_\lambda u_{\lambda_2}(p_2,m_2),\\ 
Z^{\lambda\lambda '}_{\lambda_1,\lambda_2, \lambda_3,\lambda_4}(p_1,p_2,p_3,p_4)&=&\left[\bar{u}_{\lambda_1}(p_1,m_1)\gamma^\mu P_\lambda u_{\lambda_2}(p_2,m_2)\right] \\
 & \times &\left[\bar{u}_{\lambda_3}(p_3,m_3)\gamma_\mu P_{\lambda '} u_{\lambda_1}(p_4,m_4)\right],\\
\end{array}
\end{eqnarray*}
where $P_\lambda$ stands for one of the two chiral 
projectors $P_L$ or $P_R$ and $u_{\lambda}(p,m)$ denotes the
positive energy spinor solution of the Dirac equation for a
particle  of helicity $\lambda$, four momentum $p$ and mass $m$.
The decomposition of the bispinors $u_{\lambda}(p,m)$ in terms
of the massless helicity eigenstates $\omega_{\lambda}(k)$ 
yields analytic and easy to handle expressions for the
$B$ and $Z$ functions. The amplitude is then factorized in
terms of these basic building blocks;  this fact permits compact
and transparent coding and speed of calculation.
The masses are not neglected in any stage of the calculation. 
For gaugino productions and decay, we use the \textit{widthless
approximation}. Let us illustrate it with, for instance, the
calculation of the cross section associated to
$e^+e^-\to\tilde{\chi}^0_2\tilde{\chi}^0_1\to \tilde{\chi}^0_1
\tilde{\chi}^0_1 e^+ e^- $. The total amplitude associated to a given helicity configuration of the different particles is  approximated by the product of the amplitude associated to the production of the two neutralinos ($\tilde{\chi}^0_2\tilde{\chi}^0_1$) with the amplitude corresponding to the decay of the next to lightest neutralino ($\tilde{\chi}^0_2$). The propagator squared of $\tilde{\chi}^0_2$ is approximated by a factor given by $8\pi^4/(m_{\tilde{\chi}^0_2}\Gamma_{\tilde{\chi}^0_2})$. The total amplitude expression is given by:
\begin{eqnarray*}
{\mathcal{M}} \sim \frac{2\sqrt{2}\pi^2}{m^{1/2}_{\tilde{\chi}^0_2}\Gamma^{1/2}_{\tilde{\chi}^0_2}} \sum_{\lambda_{\tilde{\chi}^0_2}}{\mathcal{M}}(e^+e^-\to\tilde{\chi}^0_2\tilde{\chi}^0_1)\times{\mathcal{M}}(\tilde{\chi}^0_2\to \tilde{\chi}^0_1 e^-e^+)
\end{eqnarray*}
The phase space integration is done through multichannel method \cite{pittau}. The left side of figure \ref{fig:beampolcorr}, illustrates the effect of
spin-correlations in angular distributions of decay electrons for 
the neutralinos case \cite{spincorrtalks}. 
Recent studies \cite{Nojiri:stau} have shown that $\tau$
polarization effects yield valuable information for the 
MSSM parameters e.g for  $\tan\beta$, the nature of the LSP 
and the mixing angle $\theta_{\tilde{\tau}}$. That is why, 
we interfaced \texttt{SUSYGEN3} to \texttt{TAUOLA} \cite{was:tauola}
which  takes into account the $\tau$ polarization accurately
in the decays.
The large effects one expects, for the momenta of the decay 
distributions are shown on the right side of figure \ref{fig:beampolcorr}.

\section{Including phases in supersymmetry searches}
\begin{figure}[h!]
\vskip -0.5cm
\dofig{2.5in}{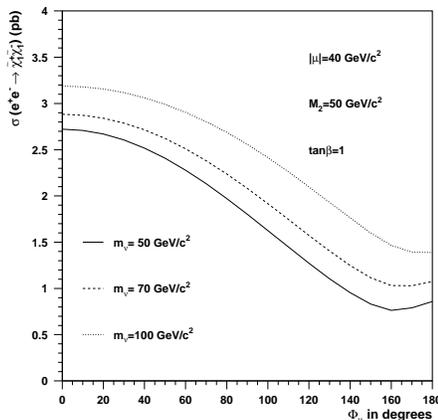}
\vskip -0.5cm
\caption{$e^+e^-\to \tilde{\chi}^+_1 \tilde{\chi}^-_1$ cross section evolution in terms of $\phi_\mu$, the phase associated to the higgsino mixing  mass parameter $\mu$. We also assume different hypotheses concerning the sneutrino $\tilde{\nu_e}$ mass \label{fig:phases}}
\end{figure}

In the MSSM, there are new potential sources of CP non 
conservation\cite{poko:phases}. Complex CP violating phases can
 arise from several parameters present in the MSSM Lagrangian: the
higgs mixing mass parameter $\mu$, the gauginos
masses $M_i$, the trilinear couplings $A_i$. 
Experimental constraints on these CP
violating phases come from the electric dipole moment 
of the electron and the neutron.
Since all the couplings 
\cite{rosi:coupling} and the different mass parameters $\mu$,
$M_1$, and the trilinear couplings $A_\tau$, $A_t$ and $A_b$ 
have been assumed to be complex by default, the introduction of
phases in the gaugino and sfermion sector for masses as well for
cross sections has been straightforward in \texttt{SUSYGEN3}.
Figure \ref{fig:phases} 
shows the chargino pair production cross section variations in
terms of  $\phi_\mu$, the phase associated to the $\mu$ parameter, 
for several values of the sneutrino mass $m_{\tilde{\nu}_e}$ and
for a value of $\tan\beta$ equal to 1. One sees that there is a 
local minimum between the two extreme values 
($\Phi_{\mu}=0,180$\ degrees), by default examined at LEP searches,
but also that it is not so deep as to raise doubts on the exhaustiveness
of "phaseless" searches. On the contrary at the linear collider,
one cannot neglect this strong dependence of the cross section
from phases.

\section{Future projects}
 Native C++ event generators are needed 
 to optimize the couplings with linear collider simulation based on
 \texttt{GEANT 4} and analysis tools. This will enable to share
 code and reuse classes between different generators.
The helicity amplitude method and 
the two basic functions, namely the $B$ and the $Z$ functions, used
to upgrade \texttt{SUSYGEN} seem to be optimally suited 
for object oriented implementation. Moreover, \textit{Multichannel
integration} looks well adapted for object treatment.

\end{document}